# Crosstalk Reduction for Superconducting Microwave Resonator Arrays

Omid Noroozian, *Student Member, IEEE*, Peter K. Day, Byeong Ho Eom, Henry G. Leduc, and Jonas Zmuidzinas, *Member, IEEE*

*Abstract*— **Large-scale arrays of Microwave Kinetic Inductance Detectors (MKIDs) are attractive candidates for use in imaging instruments for next generation submillimeter-wave telescopes such as CCAT. We have designed and fabricated tightly packed ~250-pixel MKID arrays using lumped-element resonators etched from a thin layer of superconducting TiN$_x$ deposited on a silicon substrate. The high pixel packing density in our initial design resulted in large microwave crosstalk due to electromagnetic coupling between the resonators. Our second design eliminates this problem by adding a grounding shield and using a double-wound geometry for the meander inductor to allow conductors with opposite polarity to be in close proximity. In addition, the resonator frequencies are distributed in a checkerboard pattern across the array. We present details for the two resonator and array designs and describe a circuit model for the full array that predicts the distribution of resonator frequencies and the crosstalk level. We also show results from a new experimental technique that conveniently measures crosstalk without the need for an optical setup. Our results reveal an improvement in crosstalk from 57% in the initial design down to ≤ 2% in the second design. The general procedure and design guidelines in this work are applicable to future large arrays employing microwave resonators.**

## I. INTRODUCTION AND BACKGROUND

SUPERCONDUCTING detectors are of strong interest for a number of astrophysics applications in millimeter, submillimeter, and far-infrared wavelength bands [1]. Next generation ground-based submillimeter-wave telescopes, such

Manuscript received October 16, 2011. This work was supported in part by NASA grants NNG06GC71G and NNX10AC83G, Jet Propulsion Laboratory (JPL), the Gordon and Betty Moore Foundation, and the Keck Institute for Space Studies. The devices used in this work were fabricated at the JPL Microdevices Laboratory.

O. Noroozian is with the Department of Electrical Engineering, and with the Submillimeter Astronomy research group, California Institute of Technology, Pasadena, CA 91125 USA (e-mail: omid@caltech.edu; phone: 626-376-8493).

P. Day and H.G. Leduc are with the Jet Propulsion Laboratory, Pasadena, CA 91109 USA (e-mail: peter.k.day@jpl.nasa.gov; henry.g.leduc@jpl.nasa.gov).

B. H. Eom is with the Submillimeter Astronomy research group, California Institute of Technology, Pasadena, CA 91125 USA (email: ebh@caltech.edu).

J. Zmuidzinas is with the Department of Physics, Mathematics and Astronomy, California Institute of Technology, Pasadena, CA 91125 USA, and also with the Jet Propulsion Laboratory, Pasadena, CA 91109 USA (e-mail: jonas@caltech.edu).

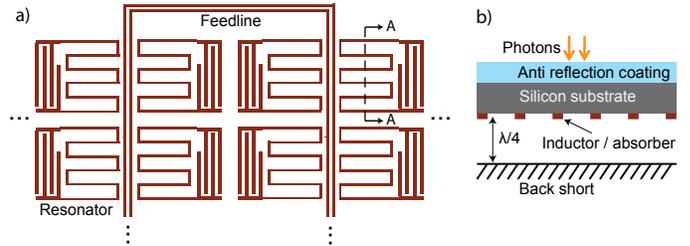

Fig. 1. (a) Schematic illustration of a 2x8 section of the geometry of a close-packed resonator array, with dark regions representing TiN$_x$ metallization. (b) Cross-sectional view along A-A in (a) of a resonator showing the illumination mechanism and the metal back short.

as CCAT [2] or future space telescopes, require focal planes with large-scale detector arrays of $10^4$–$10^6$ pixels. Submillimeter and far-infrared bolometric arrays have been growing exponentially in a "Moore's law" fashion, doubling in size every 20 months [2], and they have reached pixel counts as high as $10^4$ in the SCUBA II instrument [3]. However, further progress has been hampered by complicated and costly fabrication and readout electronics, especially the need for complex cryogenic SQUID-based multiplexing circuits. This has driven the need for simplified alternative detector designs suitable for high packing densities and with lower cost.

Arrays using Microwave Kinetic Inductance Detectors (MKID) [4-5] offer a potential solution. Figure 1(a) shows a schematic illustration of a section of an MKID array described in this work. These arrays can be easily made with a single layer of superconducting metal film deposited on a silicon substrate and pattered using conventional optical lithography. Furthermore, MKIDs are inherently multiplexable in the frequency domain, allowing ~ $10^3$ detectors to be read out using a single coaxial transmission line and cryogenic amplifier, drastically reducing cost and complexity.

An MKID uses the change in the microwave surface impedance of a superconducting thin-film microresonator to detect photons. Absorption of photons in the superconductor breaks Cooper pairs into quasiparticles, producing a change in the complex conductivity and therefore the surface impedance, which results in a perturbation of resonator frequency and quality factor. For excitation and readout, the resonator is weakly coupled to a transmission-line called a feedline. The complex amplitude of a microwave probe signal tuned on-resonance and transmitted on the feedline past the resonator is perturbed as photons are absorbed in the superconductor. The perturbation can be detected using a cryogenic SiGe transistor or InP high electron mobility transistor (HEMT) amplifier at the detector output and subsequent homodyne mixing at room temperature [4]. In an array of MKIDs, all the resonators are coupled to a shared feedline and are tuned to slightly different frequencies. They can be read out simultaneously using a comb of frequencies generated and measured using digital techniques [6, 7].

MKID arrays are being developed for astronomy at a wide range of wavelengths from millimeter waves to X-rays [7, 8, 9, 10]. Other applications of superconducting resonators are in quantum computation experiments [11, 12, 13], multiplexed readout of transition edge sensor bolometers [14], and parametric amplifiers [15]. Our previous work on mm/submillimeter-wave detection using MKIDs resulted in the construction and demonstration of a nearly complete imaging instrument called MUSIC (Multiwavelength Submillimeter kinetic Inductance Camera) [7]. The focal plane has 576 spatial pixels. For each pixel, radiation is coupled onto the chip using a planar multi-slot antenna and a low-loss superconducting Nb microstrip feed network. Superconducting lithographic band-pass filters split the mm/submillimeter signal from every antenna into 4 bands (850, 1100, 1300, and 2000 microns), and the radiation is then detected using one MKID resonator per band. Each resonator is a hybrid of an interdigitated capacitor and a coplanar waveguide (CPW) inductor mainly made from Nb, except for a short Al section at the end of the CPW. The coupled radiation is absorbed in the Al section of the resonators which has a lower superconducting energy gap than the incident mm/submillimeter radiation energy. Fabrication of the array requires $\sim$ 6 levels of lithography. In June 2010 a demonstration run of this instrument resulted in encouraging results with sensitivities reaching close to the background noise limit [8].

The arrays described in the present work are very different from the MUSIC arrays. Since they are intended for operation at a free-space wavelength of $\sim$ 350 microns (860 GHz), Nb multi-slot antennas and microstriplines cannot be used as a radiation coupling mechanism due to the resistive loss above the Nb superconducting energy gap ($\sim$ 700 GHz). To demonstrate the feasibility of large arrays of submillimeter-wave MKIDs at 350 microns, we fabricated arrays with $\sim$250 resonators using variations of Cardiff-style lumped-element resonators [16, 17]. The resonator structures were designed to act as direct absorbers of radiation, taking advantage of highly resistive $TiN_x$ films [18] to achieve a good impedance match to the incoming radiation. The $TiN_x$ film is deposited on a high-resistivity crystalline silicon substrate. The array is back-illuminated; the photons travel through the substrate and are absorbed in the $TiN_x$ resonators on the back side (Fig. 1(b)). Fresnel reflection of radiation entering the silicon substrate can be eliminated by using an anti-reflection layer. We demonstrated the basic operation of the first-generation arrays by measuring the response to a $\lambda$=215 $\mu$m bandpass-filtered blackbody source, with the results indicating $\sim$70% absorption efficiency (single polarization), comparable to the front/back power division ratio expected for silicon. It should be possible to reach 100% by using a metal backshort [19].

Although the optical response measurements for our initial design (A; see Fig. 2) were encouraging, the electromagnetic coupling between resonators combined with the high packing density (in real space and frequency space) resulted in large microwave crosstalk. We initially detected this problem through observation of non-uniform resonance frequency spacings and very large variations in resonance quality factors across our arrays. Similar effects had also been reported for arrays developed at Cardiff [19], but no detailed analysis or effective solution was available. Here we present a detailed circuit model, identify the source of the crosstalk, present a measurement technique for quantifying crosstalk, and finally present an improved resonator and array design which shows negligible crosstalk.

In Section II we explain details of our initial resonator and array design (A), and our new design (B). We identify the cause for the high inter-pixel coupling as being due to the large dipole moment of each resonator interacting with nearby resonators. To reduce the dipole moments, we modified the resonator geometry in design B so that sections with opposite charge densities and currents are close together. As another precaution, we added a grounding shield around each resonator. In Section III we briefly review the fabrication procedure. In Section IV we describe a simple model for coupling between two resonators and confirm it using electromagnetic simulations. In Section V we construct circuit models for full array arrays of type A and B. We then compare the model predictions for the mode frequencies to network analyzer measurements. In Section VI we describe a method for measuring crosstalk in the lab using a simple 'pump-probe' technique and present measurement results for both arrays. These results show that crosstalk is high ($\sim$ 57%) for design A and is dramatically reduced due to modifications in design B to $\leq$ 2%.

## II. RESONATOR AND ARRAY DESIGNS

Our arrays use lumped element resonators that are designed to efficiently absorb submillimeter-wave radiation. The resonators are made from a thin layer of superconducting titanium nitride ($TiN_x$ [18]) which has recently been shown to have several advantages over more conventional materials such as aluminum: 1) The high surface resistance of $TiN_x$ films makes it easy to design good far-IR absorbers; 2) $TiN_x$ films also have a large surface inductance that greatly increases the responsivity to photo-generated quasiparticles; 3) The high kinetic inductance also reduces the resonance frequency, thereby increasing the multiplexing density; 4) The ultra-low microwave loss in the material enables extremely high quality factors as high as $3 \times 10^7$ [18]; 5) The transition temperature is tunable over a wide range ($0 < T_c < 5$ K) by changing the nitrogen content, which allows for optimization of the detector response over a wide range of loading conditions. For these reasons $TiN_x$ is an excellent choice for our detectors.

Our initial resonator and array design (A) showed good optical absorption properties but suffered from large crosstalk between adjacent resonators. To solve this problem we created a new design (B). The two designs are explained below.

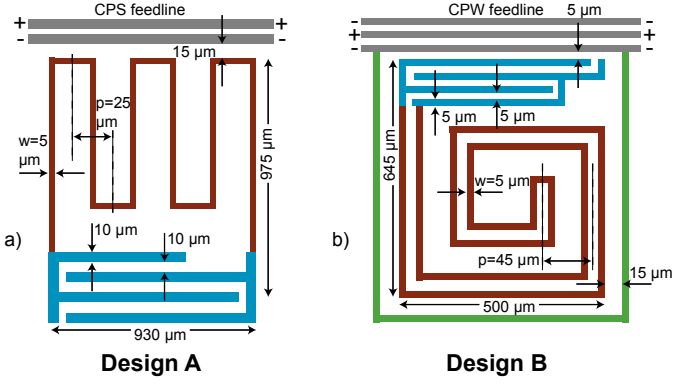

Fig. 2. Schematic of resonator designs A and B. Dimensions are not to scale and the number of meanders in design A and turns in the spiral in design B have been reduced for better visibility. The inductors are colored in red, the capacitors in blue, the feedlines in gray, and the ground shield in green.

### A. Design "A"

This design closely follows the original Cardiff proposal [16]: we use a lumped element resonator with a meandered single-line inductor and an interdigitated capacitor (IDC). A schematic of the resonator is shown in Fig. 2(a) and a photograph in Fig. 3(c). The whole resonator array is made from a single layer of ~40 nm thick TiN film with transition temperature $T_c = 4.1$ K, sheet (kinetic) inductance $L_s \approx 6.9$ pH, and sheet resistance $R_s \approx 20 \Omega$.

The inductor is the photosensitive portion of the resonator since the microwave current is large, whereas the capacitor electrodes have much lower current and therefore this portion of the structure is essentially "blind". The inductor consists of thirty-two ~1 mm long strips with a width of $w = 5$ μm and a spacing of $p = 25$ μm. The total inductance is $L \approx 60$ nH at 1.5 GHz, and kinetic inductance fraction $\alpha_{sc} = 0.74$. The kinetic inductance fraction is defined as the ratio of kinetic inductance to total inductance of the line. The total size of the pixel is ~ 1 mm × 1 mm. We used an initial estimate for $R_s$ to select the fill factor $w/p$ needed to achieve efficient submillimeter absorption. The relevant impedance-matching condition is given by

$$R_s \times p/w \approx \frac{377 \ \Omega}{1 + \sqrt{\epsilon_{si}}} \ , \qquad (1)$$

where $\epsilon_{si} \approx 11.4$ is the dielectric constant of silicon in the far-infrared at 1.5 K [20].

The capacitor has four 0.9 mm × 10 μm fingers with relatively large 10 μm gaps to reduce two-level system noise and dissipation [21]. The capacitor area is kept to ~10% of the total area in order to minimize dead space. The length of the capacitor fingers can be varied to tune the resonance frequencies in the array. At maximum finger length, the capacitance is $C \approx 0.2$ pF. The coupling to the feedline was designed for a coupling quality factor $Q_c \approx 1.7 \times 10^6$.

We fabricated a 16x14 array of these resonators. Readout is accomplished using a single coplanar strip (CPS) feedline with characteristic impedance $Z \approx 141 \ \Omega$. The physical gap between pixels is ~ 65 μm in both directions. The resonator frequencies were designed to be separated by 1.3 MHz. The lowest frequency resonator is at the top left of the array and the 224 frequencies increase linearly across the columns from top to bottom. The feedline runs across the resonators vertically and connects to the input and output side SMA connectors using circuit board transitions and wirebonds.

### B. Design "B"

A diagram of the lumped-element resonator used for design B is shown in Fig. 2(b) and a photograph is shown Fig. 3(d). The TiN film is ~20 nm thick with $T_c = 3.6$ K, $L_s \approx 11.7$ pH, and $R_s \approx 30 \ \Omega$. The inductor is a coplanar strip (CPS) with $w = 5$ μm strips and $g = 10$ μm gap. The spacing between each CPS pair is $p = 45$ μm. The inductor has a total inductance of $L \approx 43$ pH at 1.5 GHz, and $\alpha_{sc} = 0.91$. The total size of the pixel is ~ 0.65 mm × 0.5 mm. The ratio $p/w$ has been tuned using (1), where we have used an effective width $w_{eff} = 2w$ instead of $w$. The spiral shape was chosen to allow absorption in both polarizations [22]. The efficiency of the absorber has been simulated and measured and will be presented in a separate report.

The capacitor has fourteen 0.5 mm × 5 μm fingers with 5 μm gaps. The increased number of fingers allows for wider frequency tunability and better current uniformity in the inductor. At maximum finger length, the capacitance is $C \approx 0.38$ pF. The coupling to the feedline was designed for $Q_c \approx 3.8 \times 10^5$.

We fabricated a 16x16 array of these resonators (Fig. 3 (b)) and used a finite-ground coplanar waveguide (CPW) feedline (Z=115 Ω) as opposed to a CPS feedline. We added periodic grounding straps connecting CPW ground strips to eliminate the unwanted coupled slotline mode [23]. We used Nb instead of TiN for the CPW centerline in order to reduce the impedance mismatch to the 50 Ω connections, helping to reduce the observed $Q_c$ variability across the array. The physical gap spacing between pixels is ~ 35 μm in both directions. To further reduce crosstalk, the resonator frequencies are split into two groups of 128: a high frequency band (H) and low frequency band (L) that are separated by 100 MHz, and are distributed in a checkerboard pattern in the array. The frequency spacing between resonances was designed to be 1.25 MHz and 2.2 MHz in the L and H bands respectively. Starting from the top left and going down in the first column, we have resonator L1 (the lowest frequency in the L band), H1 (lowest in the H band), L2, H2, …, L8, H8. In the second column we have L9, H9, …etc. This pattern distributes the resonators in a way that keeps resonators that are close in frequency farther apart physically, reducing the pixel-pixel crosstalk.

As will be shown in Sections V and VI, design B has considerably lower crosstalk than design A. In design B the use of a double-wound (CPS) inductor places conductors with opposite polarities in close physical proximity, resulting in a good degree of cancelation of the resonator's electromagnetic fields. This confines the fields closer to the structure, reducing stray interactions between nearby resonators in a close-packed array. The grounding shield around each resonator further helps in confining the fields. As a positive side effect of the proximity of opposite polarity conductors, the geometrical inductance in design B is reduced resulting in a larger $\alpha_{sc}$. A theoretical circuit model and experimental crosstalk results for design A and B are discussed in Sections IV, V, and VI.

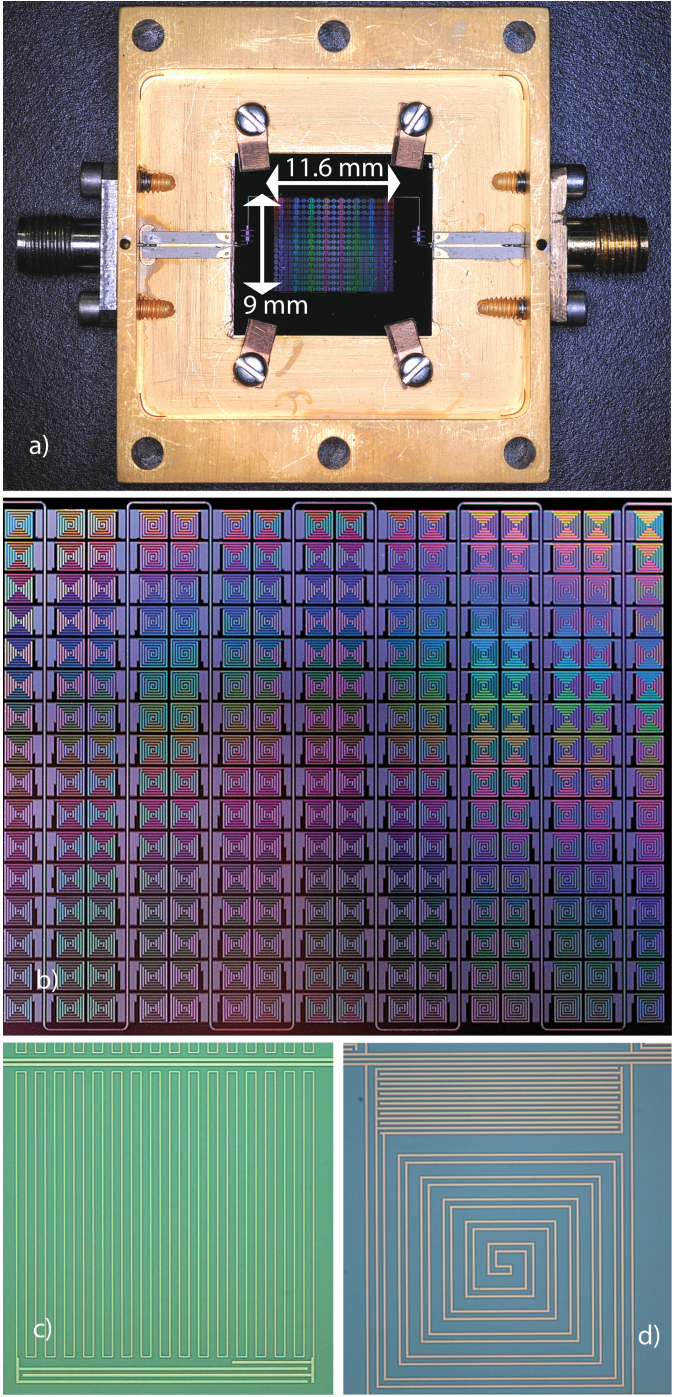

Fig. 3. (a) Device mounting box (gold plated copper) showing microstrip-CPW circuit board transitions, SMA coaxial connectors, four retaining clips, and the detector array. (b) A photograph of an earlier version of a 16x16 array of design B pixels but with no resonator ground shield and a CPS feedline instead of a CPW feedline. (c) Photograph of design A resonator. (d) Photograph of design B resonator.

## III. FABRICATION PROCEDURE

First, a layer of $TiN_x$ film is sputtered onto an ambient-temperature, high resistivity ($> 10$ k$\Omega$ cm) $\langle 100 \rangle$ silicon substrate. The substrate is cleaned with hydrofluoric acid (HF) prior to deposition. The critical temperature $T_c$ of the film is sensitive to the stoichiometry of the film and is controlled by the flow rate of the sputtering gases Ar and $N_2$ [18]. UV projection lithography is used followed by inductively coupled plasma etching using $BCl_3/Cl_2$ to pattern the resonator structures and the CPS feedline, all in one layer, for design A.

For design B the feedline is a CPW line with periodic $TiN_x$ ground straps spaced 500 μm apart. To avoid shorts caused by the straps, the centerline is initially not patterned and a 200 nm thick insulating layer of $SiO_2$ is deposited on top using RF magnetron sputtering from a high purity fused silica target. A thin layer of niobium is then deposited using DC magnetron sputtering, and is patterned using an inductively coupled plasma etcher and a mixture of $CCl_2F_2$, $CF_4$, and $O_2$, to create the centerline of the CPW feedline. The $SiO_2$ layer is then patterned using a buffered oxide etch (BOE) to remove $SiO_2$ from over the resonators.

## IV. COUPLED-RESONATORS MODEL

A simple circuit can be used to model electromagnetic coupling between two adjacent resonators in our arrays. The coupling can be capacitive, inductive, or a combination of the two. For purposes of discussion we assume a net capacitive coupling. Fig. 4(a) shows two resonators coupled with a cross-coupling capacitor $C_{12}$, with $C_1$ and $C_2$ being the capacitances of the interdigitated capacitors (IDC) and $C_1, C_2 \gg C_{12}$. (If the coupling were inductive, $C_{12}$ would be replaced by an inductor $L_{12}$ where $L_{12} \gg L$ and a similar analysis as explained below would follow). The inductor $L$ represents inductance of the meandered lines in design A and the spiral in design B resonators. It is easy to see that if the two resonators are uncoupled ($C_{12} = 0$), the two natural resonance modes are $f_{0n} = 1/2\pi\sqrt{LC_n}$ , $n = 1,2$. However, if $C_{12} \neq 0$, the frequencies of the two modes are

$$f_n = 1/2\pi\sqrt{L\lambda_n} \qquad (2)$$

where the capacitance eigenvalues are

$$\lambda_n = \bar{C} + C_{12} \mp \sqrt{C_{12}^2 + (\Delta C)^2/4} , \qquad (3)$$

and where $\bar{C} = (C_1 + C_2)/2$ and $\Delta C = (C_1 - C_2)$. The values for $C_1$, $C_2$, and $L$ can be extracted by simulating each component in an electromagnetic (EM) simulation software like Sonnet [24]. The value of the coupling element ($C_{12}$ or $L_{12}$) is difficult to extract from direct simulation. However, by examining the details of the above circuit, one can see that in a case where $C_1 = C_2 = C$, the eigenvectors are either $V_1 = V_2$ ("symmetric mode") or $V_1 = -V_2$ ("anti-symmetric mode").

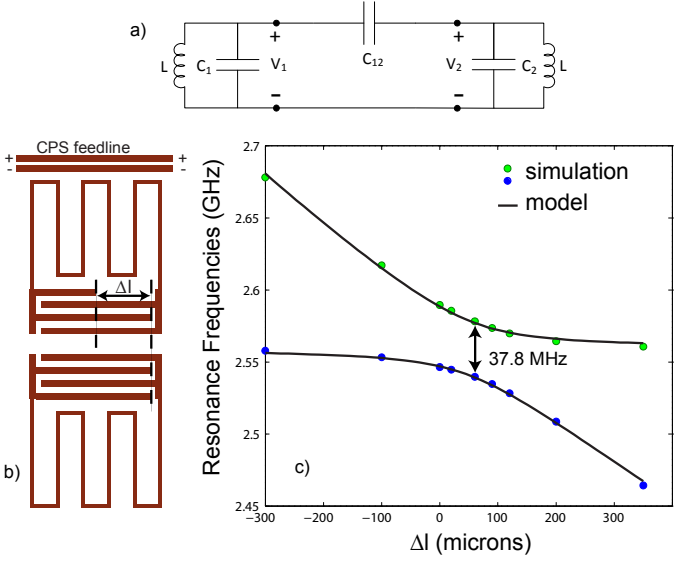

Fig. 4. (a) Circuit representation of two coupled resonators with cross-coupling capacitor $C_{12}$. (b) Schematic example of two coupled resonators where the length difference of the capacitors is indicated. Dimensions are not to scale. (c) Resonance frequencies of two coupled resonators in (b) when the finger length of one capacitor is changed show an avoided level crossing indicating a cross-coupling strength of $\delta f_{split}$=37.8 MHz. The circles are simulation results from Sonnet and the lines are a fit to (2).

We define the splitting frequency $\delta f_{split} = f_s - f_A$ and $\overline{f} = (f_s + f_A)/2$ where $f_s$ and $f_A$ refer to the symmetric and anti-symmetric mode frequencies. Depending on the type of cross-coupling, $\delta f_{split} > 0$ for capacitive coupling (as in Fig. 4) and $\delta f_{split} < 0$ for inductive coupling. From this, it is easy to show that

$$C_{12} \approx \left(\frac{\delta f_{split}}{\overline{f}}\right) C \quad , \text{ capacitive coupling} \qquad (4)$$

$$L_{12} \approx -\left(\frac{\overline{f}}{\delta f_{split}}\right) L \quad , \text{ inductive coupling.} \qquad (5)$$

These equations allow us to determine the value of the coupling element assuming that we know $\delta f_{split}$. The quantity $|\delta f_{split}|$ is a measure of the coupling strength and increases as the coupling gets stronger.

We used Sonnet software [24] to directly simulate our coupled resonators to extract $\delta f_{split}$. Fig. 4(b) shows the schematic of two such coupled resonators. We run multiple simulations, each time slightly changing the capacitance value of one resonator (using its IDC finger length) and keeping the other capacitance constant. The resulting frequencies are shown in Fig. 4(c) (circles) where the horizontal axis is proportional to capacitance difference. As the difference in capacitance approaches zero, an avoided crossing appears. Our circuit model agrees well with this behavior as is evident from the solid lines in Fig. 4(c), which are from a fit to (2). The minimum separation in the two curves is equal to $\delta f_{split}$ and can be used in (4) or (5) to estimate the value of the coupling element.

## TABLE I
## COUPLING SPLITTING FREQUENCIES

| Design A | | Design B | |
|---|---|---|---|
| Configuration | $\delta f_{split}$ (MHz) | Configuration | $\delta f_{split}$ (MHz) |
| | 36.2 | | 0.20 |
| | 60.4 | | 1.75 |
| | 8.6 | | 0.25 |
| | 28.2 | | 1.18 |
| | 6.9 | | 0.35 |

It should be mentioned that identifying whether the cross-coupling mechanism is capacitive or inductive can be achieved through simulations by noting that the symmetric mode will generally have a stronger coupling to the feedline (lower $Q_c$) compared to the anti-symmetric mode, since the currents injected in the feedline will be in-phase as opposed to 180° out-of-phase. This observation yields the sign of $\delta f_{split}$.

## V. FULL ARRAY CIRCUIT MODEL AND SIMULATION

In order to predict the behavior of a complete array of size $N$, we calculated the eigenfrequencies $\omega$ for an equivalent circuit. The circuit consists of identical inductors attached to the $N$ ports of a capacitance network that takes into account all of the nearest neighbor resonator couplings using the actual positions of the resonators with respect to each other and with respect to the feedline:

$$\frac{1}{L\omega^2}\begin{pmatrix} V_1 \\ \vdots \\ V_N \end{pmatrix}$$

$$= \begin{pmatrix} C_1 + \sum_{j\neq 1} C_{1j} & -C_{12} & \cdots & -C_{1N} \\ -C_{21} & C_2 + \sum_{j\neq 2} C_{2j} & & \vdots \\ \vdots & & \ddots & \\ -C_{N1} & \cdots & & C_N + \sum_{j\neq N} C_{Nj} \end{pmatrix} \begin{pmatrix} V_1 \\ \vdots \\ V_N \end{pmatrix} \qquad (6)$$

where $C_{ij}$ ($i \neq j$) are the cross coupling capacitances, $C_i$ are capacitances of the IDCs, and $L$ is the meander or spiral inductance. The values for the $C_{ij}$ were calculated by first simulating all the nearest-neighbor two-resonator configurations in each array in Sonnet and extracting their corresponding splitting frequencies $\delta f_{split}$. These are listed in

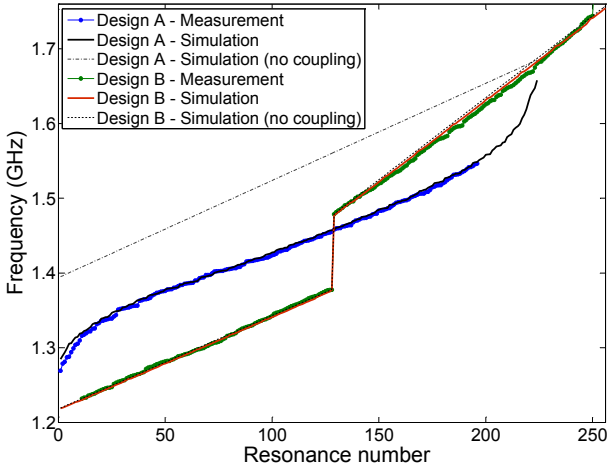

Fig. 5. Measurements of array resonance frequencies for both designs A and B and comparison to simulation. The two "simulation (no coupling)" curves are for when $C_{ij}$ ($i \neq j$) = 0.

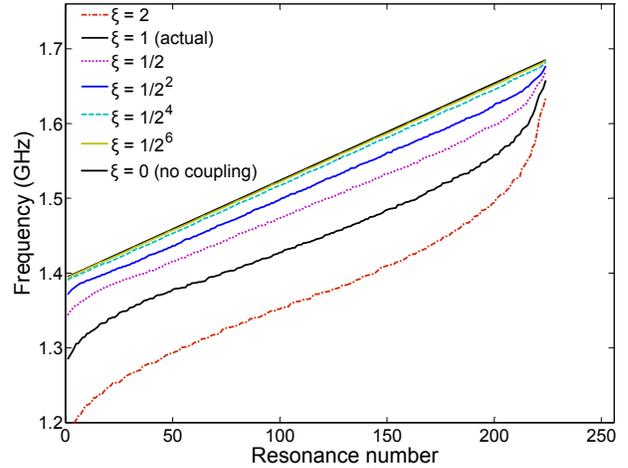

Fig. 6. A series of simulations showing how the frequency curve shape evolves as the coupling strength is varied by tuning the scaling factor $\xi$.

Table I. One can already see that the splitting frequencies for design B are considerably smaller than design A and also smaller than the frequency spacing between resonances. We then used (4) together with Table I to convert $\delta f_{split}$ values into corresponding $C_{ij}$'s. The resulting array eigenfrequencies from the simulations are shown in Fig. 5 for both designs, and are compared to network analyzer measurements. The measurement setup was the same as described later in Fig. 9 except with the pump tone turned off. Assumptions that were used in the simulation are explained in Section II. As can be seen, the simulations are in excellent agreement with measurements for both arrays, confirming our circuit model for coupled resonator arrays.

The specific inverted-S shape of the curve for design A is a characteristic feature of highly coupled arrays. This has been further explored in Fig. 6 where we have plotted the eigenfrequencies for design A for a range of coupling strengths by scaling the values in Table I by the same factor. As can be seen, the S-shape gradually disappears as we reduce the strength of the coupling, and becomes negligible by the time it is down by a factor of 16. The same S-shape eventually also appears in both bands of array B when we artificially scale up the coupling strength, but for the actual array the effect is negligible thanks to the much smaller splitting frequencies and the checkerboard frequency scheme.

Equation (6) also yields the normalized eigenvectors for the resonance modes. An example is shown visually in Fig. 7 where it can be seen that the mode is highly delocalized for design A whereas it is highly localized for design B. The colors show the amount of normalized energy contributed by each physical resonator to the resonance mode. The energy in each of the four diagonal nearest neighbor pixels in design B is more than 40 dB lower than the main pixel. These results strongly indicate that crosstalk is dramatically reduced in design B. This conclusion is confirmed by direct measurements and simulations of the crosstalk, which we present next.

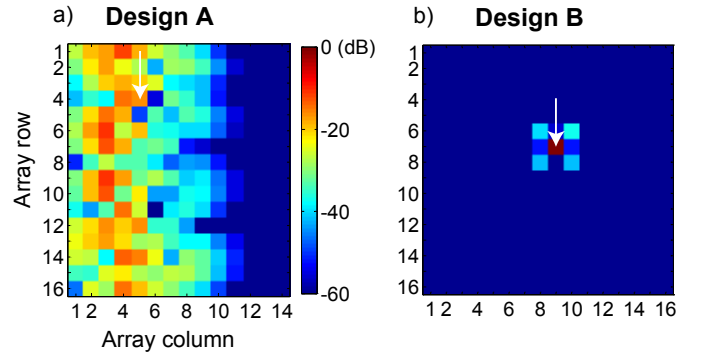

Fig. 7. Normalized energy ($20 \times \log_{10} V_n$) in voltages $V_1$ to $V_N$ across the array for a specific mode number (#68) in both arrays is shown in color. In an uncoupled array, this mode number would purely correspond to the resonator in position # 68 in array A, and in position # 135 in array B (indicated by arrows). However, due to strong coupling in array A, energy is distributed over many resonators, while in array B the energy is well localized.

## VI. CROSSTALK

One method to measure crosstalk is to illuminate a single physical resonator on the array with submillimeter photons and to look for a response in other resonances. This approach is difficult because confining the far-IR light to one pixel requires a complicated optical setup. Instead, we developed a very simple "pump-probe" technique where we apply a microwave "pump" tone to a resonance and observe the response from the other resonances. This technique exploits the fact that the kinetic inductance of a superconductor generally is nonlinear [5,25] and can change as a function of the microwave current:

$$L_{kin}(I) \approx L_{kin}(0)[1 + I^2/I_*^2] , \qquad (7)$$

where $I$ is the microwave current in the inductor, and $I_*$ sets the scale of the nonlinearity and is often comparable to the DC critical current. By applying a strong microwave pump tone to one of the resonance modes $l$ with frequency $f_1^l$, the microwave currents in the inductors that participate in that mode cause the inductance values to increase slightly according to (7), so the mode frequency decreases to a new value $f_2^l$ that may be characterized by the frequency shift $\delta f^l = (f_2^l - f_1^l)$. In an array where the pixels are coupled, this will also result in shifts in other modes, and by comparing these shifts to $\delta f^l$ we can experimentally measure crosstalk for each mode. If $f_1^k$ and $f_2^k$ are the frequency of a certain "probed" mode $k$ when the pump is applied on-resonance and off-resonance respectively, then a quantitative measure of the crosstalk may be defined as $\delta f^k / \delta f^l = (f_2^k - f_1^k)/\delta f^l$. The effect of the nonlinearity-induced shifts in the inductances may be analyzed by generalizing (6) to include non-equal inductors, and by noting that $\delta f^l / f_1^l \sim 10^{-6}$ in our measurements so the use of linear perturbation theory is very well justified. The result of this calculation yields an expression for the crosstalk values

$$\chi_{kl} = \frac{\delta f^k}{\delta f^l} = \frac{f^l}{f^k} \frac{\sum_{i=1}^{N} |V_i^k|^2 |V_i^l|^2}{\sum_{i=1}^{N} |V_i^l|^4}. \qquad (8)$$

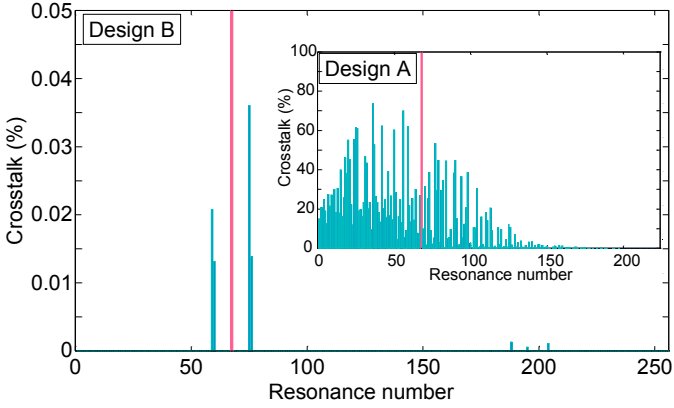

Fig. 8. Crosstalk simulations for full-size arrays A and B. In both simulations mode number 68 (indicated in red) is pumped. By definition, crosstalk for the pumped mode is 100%. Note that the scales are very different in the two plots.

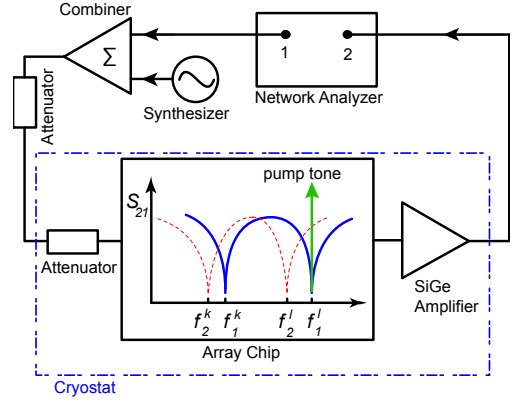

Fig. 9. Illustration of the setup for measuring the resonances and the crosstalk. The resonators are cooled down to below 100 mK in a cryogenic refrigerator, and are read out using a network analyzer. A SiGe transistor amplifier [26] at 4 K is used to amplify the signal. The synthesizer pump power is combined with readout power using a 3 dB power combiner. The pump signal frequency is tuned on a resonance (blue curve) which causes the resonance to shift (red curve). A nearby coupled resonance also shifts as a result.

Here $V_i^k$ and $V_i^l$ are the voltage eigenvector components for modes $k$ and $l$ in our circuit model (Fig. 7). This result demonstrates that modes whose "energy overlap" is large will have significant crosstalk.

Figure 8 shows crosstalk simulated for both arrays A and B using (8). In both simulations an arbitrary mode number (#68) was pumped. The simulations show that crosstalk is very high in design A (up to ∼ 75%) where many other modes are affected by the pump, while in design B there is almost no crosstalk down to a level of ∼ 0.04%.

The measurement setup is illustrated in Fig. 9 where a synthesizer provides microwave power (pump) at the frequency of one of the resonance modes ($f_1^l$). All the resonances were probed using a network analyzer in a relatively low power mode (∼ -100 dBm on the feedline), so that the pump power was dominant (∼ -80 dBm).

Crosstalk measurement results are shown in Fig. 10(a) and (b). Both plots are for a group of resonances that have frequencies not too far from the pumped resonance. Figure 10(a) clearly shows that design A is dominated by crosstalk as large as 57%. Figure 10(b) shows that by going to design B, crosstalk dramatically reduces down to a maximum of 2%. The error bars are a result of the fits to the resonances by a fitting code [27] used to fit the data from the network analyzer. Because the network analyzer scans were taken at relatively low power, higher measurement times were required which made the data susceptible to various noise sources including network analyzer frequency drift and magnetic fields affecting the resonance positions [28]. The simulations shown in Fig. 8 support the measurements and suggest that the actual crosstalk in design B could be much lower than the experimental upper limit of 2%.

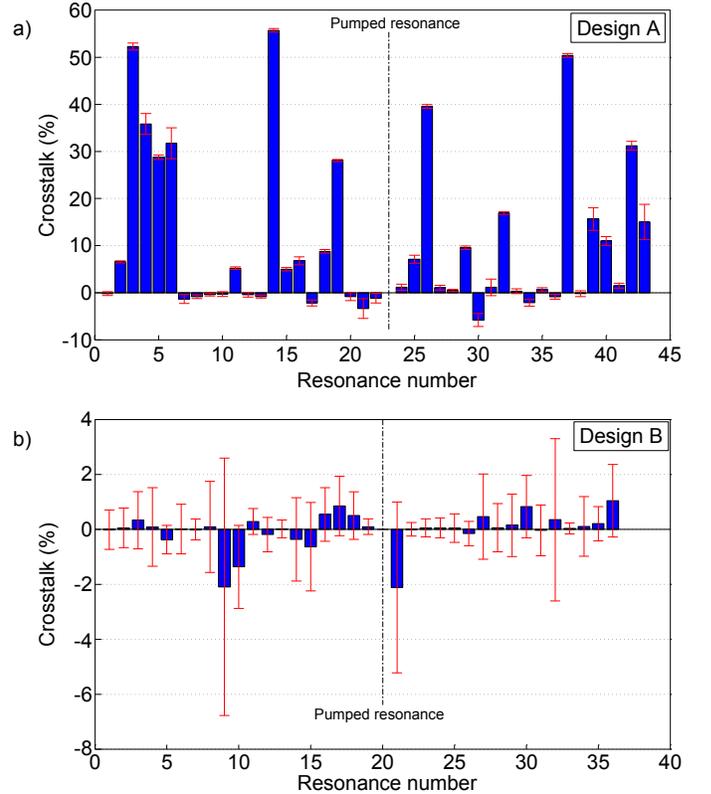

Fig. 10. Crosstalk measurement results for designs A and B. The (frequency) position of the pumped resonance is shown by the dashed line. The red bars indicate the measurement error.

## VII. Summary

We successfully fabricated ~250 pixel arrays of submillimeter-wave MKIDS using TiN on silicon. We demonstrated the basic operation of these arrays by measuring their response to a blackbody source. However, the arrays suffered from crosstalk between individual pixels due to large resonator dipole moments interacting with nearby resonators. We developed a simple and general circuit model that accounts for the crosstalk and agrees well with measurements of the array frequency curve patterns and crosstalk. We created a new design that eliminates crosstalk in two ways: 1) by reducing the dipole moment of each resonator and 2) by distributing the resonators across the array in a checkerboard pattern. We developed a simple "pump-probe" technique to directly measure crosstalk, and showed that crosstalk in design B is indeed smaller and is dramatically reduced to $\leq 2\%$. The general procedure and design guidelines in this work are applicable to future large-scale arrays of microwave resonators for telescopes like CCAT.

## Acknowledgment

The authors would like to thank Sunil Golwala, David Moore, Loren Swenson, and Ran Duan with the California Institute of Technology, for helpful discussions.

## References

[1] J. Zmuidzinas and P. L. Richards, "Superconducting detectors and mixers for millimeter and submillimeter astrophysics," *Proc. IEEE*, vol. 92, no. 10, pp. 1597-1616, Oct. 2004.

[2] Science Frontiers Panels, Program Prioritization Panels, Committee for a Decadal Survey of Astronomy and Astrophysics, "Panel Reports—New Worlds, New Horizons in Astronomy and Astrophysics," Nat. Research Council, Washington D.C., 2011 [Online]. Available: http://www.nap.edu/catalog.php?record_id=12982

[3] W. S. Holland, W. D. Duncan, B. D. Kelly, K. D. Irwin, A.J. Walton, P. A. R. Ade, and E.I. Robson, "SCUBA-2: A large format submillimetre camera on the James Clerk Maxwell Telescope," in *Proc. SPIE Millimeter and Submillimeter Detectors for Astronomy Conf.*, 2003, vol. 4855, pp. 1-18.

[4] P. K. Day, H. G. LeDuc, B. A. Mazin, A. Vayonakis, and J. Zmuidzinas, "A broadband superconducting detector suitable for use in large arrays," *Nature*, vol. 425, no. 6960, pp. 817-821, Oct. 2003.

[5] J. Zmuidzinas, "Superconducting Microresonators: Physics and Applications", *Annu. Rev. Cond. Mat. Phys*, vol. 3, pp. 15.1-15.46, 2012 (in press).

[6] B.A. Mazin, P.K. Day, K.D. Irwin, and C.D. Reintsema, "Digital readouts for large microwave low-temperature detector arrays," *Nuclear Instruments and Methods in Physics Research Section A*, vol. 599, no. 2, pp. 799–801, April 2006.

[7] P. R. Maloney, N. G. Czakon, P. K. Day, T. P. Downes, R. Duan, J. Gao, J. Glenn, S. R. Golwala, M. I. Hollister, H. G. Leduc, B. A. Mazin, O. Noroozian, H. T. Nguyen, J. Sayers, J. A. Schlaerth, S. Siegel, J. E. Vaillancourt, A. Vayonakis, P. R. Wilson, and J. Zmuidzinas, "MUSIC for sub/millimeter astrophysics," in *Proc. SPIE Millimeter, Submillimeter, and Far-Infrared Detectors and Instrumentation for Astronomy V Conf.*, 2010, vol. 7741, Art. ID 77410F.

[8] J. A. Schlaerth, J. Glenn, P. R. Maloney, N. G. Czakon, T. P. Downes, R. Duan, S. R. Golwala, M. I. Hollister, O. Noroozian, S. Siegel, A. Vayonakis, J. Zmuidzinas,P, K. Day, H. G. LeDuc, H. T. Nguyen, J. Sayers, P. R. Wilson, J. Gao, B. A. Mazin, and J. E. Vaillancourt, "MKID multicolor array status and results from DemoCam" in *Proc. SPIE Millimeter, Submillimeter, and Far-Infrared Detectors and Instrumentation for Astronomy V Conf.*, 2010, vol. 7741, Art. ID 774109.

[9] B.A. Mazin, K. O'Brien, S. McHugh, B. Bumble, D. Moore, S. Golwala, and J. Zmuidzinas, "ARCONS: A highly multiplexed superconducting optical to near-IR camera," in *Proc. SPIE Ground-based and Airborne Instrumentation for Astronomy III Conf.*, 2010, vol. 7735, Art. ID 773518.

[10] A. Monfardini, A. Benoit, A. Bideaud, L. J. Swenson, M. Roesch, F. X. Desert, S. Doyle, A. Endo, A. Cruciani, P. Ade, A. M. Baryshev, J. J. A. Baselmans, O. Bourrion, M. Calvo, P. Camus, L. Ferrari, C. Giordano, C. Hoffmann, S. Leclercq, J. F. Macias-Perez, P. Mauskopf, K. F. Schuster, C. Tucker, C. Vescovi, S.J.C. Yates, "A Dual-Band Millimeter-Wave Kinetic Inductance Camera for the IRAM 30 m Telescope," *Astrophysical Journal Supplement Series*, vol. 194, Art. ID 24, June 2011.

[11] J. B. Hertzberg, T. Rocheleau, T. Ndukum, M. Savva, A. A. Clerk, and K. C. Schwab, "Back-action-evading measurements of nanomechanical motion," *Nature Physics*, vol. 6, no. 3, pp. 213 – 217, March 2010.

[12] J. Koch, A. A. Houck, K. Le Hur, and S. M. Girvin, "Time-reversal-symmetry breaking in circuit-QED-based photon lattices," *Phys. Rev. A*, vol. 82, no. 4, Art. ID 043811, Oct. 2010.

[13] C. M. Wilson, T. Duty, M. Sandberg, F. Persson, V. Shumeiko, and P. Delsing, "Photon Generation in an Electromagnetic Cavity with a Time-Dependent Boundary," *Phys. Rev. Lett.*, vol. 105, no. 23, Art. ID 233907, Dec. 2010.

[14] J. A. B. Mates, G. C. Hilton, K. D. Irwin, L. R. Vale, and K. W. Lehnert, "Demonstration of a multiplexer of dissipationless superconducting quantum interference devices," *Appl. Phys. Lett.*, vol. 92, no. 2, Art. ID 023514, Jan. 2008.

[15] E. A. Tholen, A. Ergul, E. M. Doherty, F. M. Weber, F. Gregis, and D. B. Haviland, "Nonlinearities and parametric amplification in superconducting coplanar waveguide resonators," *Appl. Phys. Lett.*, vol. 90, no. 25, Art. ID 253509, June 2007.

[16] S. Doyle, P. Mauskopf, J. Naylon, A. Porch, and C. Duncombe, "Lumped Element Kinetic Inductance Detectors," *Journal of Low Temperature Physics*, vol. 151, pp. 530–536, April 2008.

[17] O. Noroozian, P. K. Day, Byeong Ho Eom, Henry G. Leduc, and Jonas Zmuidzinas, "Microwave crosstalk in lumped element far-IR MKIDs," in *Proc. 35th Int. Conf. Infrared Millimeter and Terahertz Waves (IRMMW-THz)*, 2010, © IEEE. doi: 10.1109/ICIMW.2010.5612440

[18] H. G. Leduc, B. Bumble, P. K. Day, A. D. Turner, B. H. Eom, S. Golwala, D. C. Moore, O. Noroozian, J. Zmuidzinas, J. Gao, B. A. Mazin, S. McHugh, and A. Merrill, "Titanium Nitride Films for Ultrasensitive Microresonator Detectors," *Appl. Phys. Lett.*, vol. 97, no. 10, Sept. 2010, Art. ID 102509.

[19] S. Doyle, P. Mauskopf, J. Zhang, S. Withington, D. Goldie, D. Glowacka, A. Monfardini, L. Swenson, and M. Roesch, "Optimisation of Lumped Element Kinetic Inductance Detectors for use in ground based mm and sub-mm arrays," *AIP Conf. Proc.*, 2009, vol .1185, pp. 156-159.

[20] James W. Lamb, "Miscellaneous data on materials for millimetre and submillimetre optics," *Int. J. Infrared and Millimeter Waves*, vol. 17, no. 12, pp. 1997-2034, Dec. 1996.

[21] O. Noroozian, J.S. Gao, J. Zmuidzinas, H.G. Leduc, B.A. Mazin, "Two-level system noise reduction for Microwave Kinetic Inductance Detectors," Proc. 13th Int. Workshop on Low Temperature Detectors, *AIP Conf. Proc.*, 2009, vol. 1185, pp. 148-151.

[22] A. Brown, W. Hsieh, S. H. Moseley, T. R. Stevenson, K. U-yen, E. J. Wollack, "Fabrication of an absorber-coupled MKID detector and readout for sub-millimeter and far-infrared astronomy," in *Proc. SPIE Millimeter, Submillimeter, and Far-Infrared Detectors and Instrumentation for Astronomy V Conf.*, 2010, vol. 7741, Art. ID 77410P.

[23] G.E. Ponchak, J. Papapolymerou, M.M. Tentzeris, "Excitation of coupled slotline mode in finite-ground CPW with unequal ground-plane widths," *IEEE Trans. Microw. Theory Tech.*, vol. 53, no. 2, pp. 713–717, Feb. 2005.

[24] Sonnet Software Inc. [Online]. Available: http://www.sonnetusa.com/

[25] T. Dahm and D. J. Scalapino, "Theory of intermodulation in a superconducting microstrip resonator," *J. Appl. Phys.*, vol. 81, no. 4, pp. 2002-2209, Feb. 1997.

[26] S. Weinreb, J. Bardin, H. Mani, and G. Jones, "Matched wideband low-noise amplifiers for radio astronomy," *Rev. Sci. Instrum.*, vol. 80, no. 4, April 2009, Art ID 044702.

[27] J. Gao, "The physics of superconducting microwave resonators," Ph.D. dissertation, Dept. Phys., California Inst. Of Tech., Pasadena, CA, 2008.

[28] J. E. Healey, T. Lindstrom, M. S. Colclough, C. M. Muirhead, and A. Y. Tzalenchuk, "Magnetic field tuning of coplanar waveguide resonators," *Appl. Phys. Lett.*, vol. 93, no. 4, Art. ID 043513, July 2008.

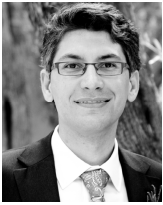

**Omid Noroozian** (S'07) received the B.S. degree in electronics engineering from Sharif University of Technology, Tehran, Iran, in 2004, the M.Sc. degree in microelectronics from Delft University of Technology, Delft, The Netherlands, in 2006, and the M.S. degree in applied physics from California Institute of Technology (Caltech), Pasadena, CA, in 2009. He is currently working toward the Ph.D. degree in Electrical Engineering at Caltech.

He was a Research Assistant with the Physics of NanoElectronics group of the Kavli Institute of Nanoscience at TUDelft from 2005 to 2006, where he worked on design and fabrication of superconducting tunnel junction heterodyne mixer chips for the Atacama Large Millimeter Array (ALMA) telescope. Since 2006, he has been a Research Assistant with the Submillimeter Wave Astrophysics group at Caltech where he is working on development of kinetic inductance detector technology for large-scale submillimeter/far infrared arrays for the Caltech Submillimeter Observatory (CSO) and the Cornell Caltech Atacama Telescope (CCAT).

**Peter K. Day**

**Byeong Ho Eom**

**Henry G. LeDuc**

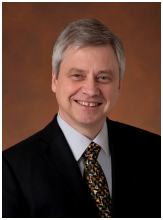

**Jonas Zmuidzinas** (Member, IEEE) received the B.S. degree in physics from the California Institute of Technology (Caltech), Pasadena, in 1981 and the Ph.D. degree in physics from the University of California, Berkeley, in 1987. He was a Postdoctoral Fellow in 1988–1989 at the University of Illinois, Urbana-Champaign. He joined the physics faculty at Caltech in 1989 and currently serves as the Merle Kingsley Professor of Physics. He also has an appointment at Caltech's Jet Propulsion Laboratory (JPL) where he serves as JPL's Chief Technologist.